\documentclass[aps,prl,twocolumn,superscriptaddress]{revtex4-1}

\usepackage{amsmath}
\usepackage{graphicx}
\usepackage{pifont}
\usepackage{array}
\usepackage{rotate}
\usepackage{vmargin}
\usepackage{lineno}
\usepackage{fancyhdr}
\usepackage{color}
\usepackage{url}
\usepackage{wasysym}

\begin{document}


\sloppy


\title{Search for magnetic monopoles in polar volcanic rocks}
\author{K.~Bendtz} 
\author{D.~Milstead}
\affiliation{Fysikum, Stockholm University, Stockholm, Sweden}
\author{H.-P.~H\"achler}
\author{A.~M.~Hirt}
\affiliation{Department of Earth Sciences, Swiss Federal Institute of Technology, Zurich, Switzerland}
\author{P.~Mermod} 
\email{philippe.mermod@cern.ch}
\affiliation{D\'epartement de Physique Nucl\'eaire et Corpusculaire, University of Geneva, Geneva, Switzerland}
\author{P.~Michael}
\affiliation{Department of Geosciences, University of Tulsa, Tulsa, USA}
\author{T.~Sloan}
\affiliation{Department of Physics, Lancaster University, Lancaster, United Kingdom}
\author{C.~Tegner}
\affiliation{Department of Geoscience, Aarhus University, Aarhus, Denmark}
\author{S.~B.~Thorarinsson} 
\affiliation{Nordic Volcanological Center, Institute of Earth Sciences, University of Iceland, Reykjav\`ik, Iceland}


\begin{abstract}
For a broad range of values of magnetic monopole mass and charge, the abundance of monopoles trapped inside the Earth would be expected to be enhanced in the mantle beneath the geomagnetic poles. A search for magnetic monopoles was conducted using the signature of an induced persistent current following the passage of igneous rock samples through a SQUID-based magnetometer. A total of 24.6~kg of rocks from various selected sites, among which 23.4~kg are mantle-derived rocks from the Arctic and Antarctic areas, was analysed. No monopoles were found and a 90\% confidence level upper limit of $9.8\cdot 10^{-5}$/gram is set on the monopole density in the search samples.
\end{abstract}

\maketitle

The existence of magnetic monopoles was postulated in 1931 by Dirac as a means to explain electric charge quantisation~\cite{Dirac1931,Dirac1948}. The Dirac quantisation argument predicts that the fundamental magnetic charge $q_m=gec$ (in this definition $q_m$ is in SI units and $g$ is a dimensionless quantity) is a multiple of the Dirac charge: $g=Ng_D$ with $g_D=68.5$ and $N$ an integer number. Magnetic monopoles are also fundamental ingredients in grand unification theories~\cite{Hooft1974}. Although grand unification monopoles would typically have masses of the order of the unification scale ($m\sim 10^{16}$~GeV), there are generally no tight theoretical constraints on the mass of a monopole. 

Calculations within nonrelativistic quantum theory indicate that monopoles would bind to non-zero-spin nuclei through magnetic moment coupling, with binding energies of the order of several hundred keV when assuming a hard core~\cite{Milton2006}. Such binding is assumed as a working hypothesis in the present search. If isolated monopoles exist in Nature, they are stable by virtue of magnetic charge conservation, and they either reside inside astronomical bodies or move freely through open space to form a galactic halo. Throughout this paper, ``stellar'' denotes monopoles already trapped in stardust before the formation of the Solar System, and ``cosmic'' denotes free monopoles reaching the Solar System at a later time. 

Signatures of direct monopole pair production have been explored at past high-energy particle colliders including the LEP, HERA and  Tevatron~\cite{MODAL,Pinfold:1993mq,TEVATRONSQUID2004,HERASQUID,Abulencia:2005hb,OPALdirect} and are being investigated with the Large Hadron Collider~\cite{MonoATLAS11,SMPPheno11}. However, monopoles with masses above 7~TeV cannot be produced within current collider programs. In this work, which probes monopoles in the mass range between the weak scale and the grand unification scale, it is assumed that monopoles may exist as relics produced out of thermal equilibrium in the very early Universe. Models of cosmological inflation allow relic monopoles to be diluted down to noncatastrophic abundances~\cite{Preskill1984}. However, the various inflationary scenarios which have been proposed can make very different monopole abundance predictions~\cite{Lyth99}. Other unknowns are the monopole-antimonopole annihilation cross section and the detailed mechanisms by which monopoles may have bound to matter during primordial nucleosynthesis. Even though there are presently no adequate models that describe to which extent relic monopoles would have accumulated inside astronomical bodies or be present in cosmic rays, abundances and fluxes can be constrained by experiments. Monopoles in flight have been sought with array detectors. These set tight constraints on the flux of cosmic monopoles incident on Earth~\cite{Indu84,Indu86a,Indu86b,Indu90,Indu91a,Indu91b,Ohya91,MACRO2002,SLIM2008,BAIKAL2008,RICE2008,AMANDA2010,ANITA2011,ANTARES2012} (only the most significant results are given here; see~\cite{PDG2012} for a complete list). Trapped monopoles have previously been sought in hundreds of kilograms of samples from the Earth's crust~\cite{Goto63,Fleischer69,Kolm71,Carrigan76,Kovalik86,Longo95}, in rocks from the Moon's surface~\cite{Moon71,Moon73}, and in meteorites~\cite{Goto63,Longo95}. This work presents the first search for monopoles in terrestrial igneous rocks at high latitudes.

Large planetary bodies such as the Earth were molten during their formation and this has lead to large-scale chemical differentiation. During this early phase stellar monopoles, if present, will likely have sunk to the planet's core ~\footnote{Even when using pessimistic monopole pair production cross section estimates, collider searches would have been able to produce and observe Dirac monopoles with a mass of the order of 500~GeV or lower~\cite{Abulencia:2005hb,MonoATLAS11}, which is heavier than the heaviest nuclei.}. Stellar monopoles should therefore be depleted in planetary crusts, while the deep interiors of planets and stars, as well as the insides of some meteoroids, asteroids and comets, would be the only places likely to contain them in non-negligible amounts. 

Monopoles inside astronomical bodies of low viscosity possessing stable dipole magnetic fields would move to positions along the magnetic axis where the magnetic force $F_m=q_m B$ ($B$ is the vertical component of the magnetic field) and gravitational force $F_g=ma$ ($a$ is the gravitational acceleration) are in equilibrium: 

\begin{equation}
m = \frac{g_DecB}{a}\frac{g}{g_D} = A\frac{g}{g_D}
\label{equ:forcebalance}
\end{equation}

Although the early configuration of the Earth's internal magnetic field is poorly known, paleomagnetic data suggest that the Earth possessed a dipole field since at least $\sim 3.5$ billion years~\cite{paleo1,paleo2,paleo3}. The configuration of the field close to the Earth's core may be more complex, but the simple assumption of a dipole field over geologic time is reasonable. Carrigan estimated that monopoles with $g=g_D$ and $m=10^{16}$~GeV would accumulate near the Earth's inner core, and developed a model of how monopole annihilation during geomagnetic reversals would contribute to the planet's internal heat, thus limiting the grand-unification-mass  monopole density inside the Earth to less than 
$\sim 10^{-4}$/gram~\cite{Carrigan80}. On the other hand, a lighter mass or higher magnetic charge will raise the equilibrium depth.  We consider monopoles attached to nuclei with an equilibrium position above the core-mantle boundary. Down to a depth of 2900~km, the Earth's mantle plays the role of an insulator between the molten outer core and the crust and has the properties of a plastic solid. Although mantle dynamics are complex and various competing geodynamical models exist, it can generally be assumed that the mantle slowly convects as a whole, with a full cycle taking approximately 400$-$500 million years~\cite{convection}. 
Monopoles caught in the solid mantle would be unable to move freely. Instead, monopoles of both polarities would be transported up and down along with mantle convection regardless of the field direction. 
Upon reaching the core-mantle boundary, they would sink through the liquid core due to the high mass, before being attracted in the general direction of the polar regions due to the magnetic charge. 
Over geologic time monopoles would migrate towards the magnetic axis. At the Earth's pole, $a=9.8$~m$\cdot$s$^{-2}$ and $B=6.5\cdot 10^{-5}$~T, in which case Equation~\ref{equ:forcebalance} yields $A_{surface}=1.2\cdot 10^{13}$~GeV (presently GeV is a unit of mass). A monopole carrying a single Dirac charge ($g=g_D$) and a mass of 10$^{13}$ GeV or lower would therefore be expected to be found beneath the Earth's polar crust and in melts below polar regions. A monopole carrying a multiple of the Dirac charge is allowed to possess a proportionally higher mass. This mass bound is conservative because monopoles with equilibrium anywhere inside the mantle may still reach the surface through mantle convection (the core-mantle boundary corresponds to $A_{boundary}=4\cdot 10^{14}$~GeV). In a naive model, one may assume that monopoles would be distributed randomly throughout the whole mantle depth up to a distance from the magnetic axis equal to the core radius of 3400~km (this corresponds to latitudes $> 57^\circ$), and absent everywhere else. This results in a concentration of monopoles 6 times higher in polar mantle-derived rocks than averaged over the Earth's mass. 

The samples used in this search were restricted to mantle-derived igneous rocks with negligible levels of crustal contamination, emplaced at high ($>63^\circ$) latitudes. Basaltic rocks from hotspots -- volcanic regions under which the mantle is thought to be locally hotter, causing an ascending mantle plume -- are particularly attractive as they are likely to include material from deep inside the mantle. Iceland and Hawaii are among the best known examples of hotspots for which there is evidence that the erupted material comes from more than 600~km depth and possibly as deep as the core-mantle boundary~\cite{hotspots1,hotspots2}. Other active hotspot sites at high latitudes, but for which the role of mantle plumes is debated~\cite{plume_controversy}, include Jan Mayen Island (Arctic Ocean)~\cite{JanMayen} and Ross Island (Southern Victoria Land, Antarctica)~\cite{Erebus}. Large igneous provinces (LIPs) are also of interest for this work. These massive magmatic provinces are dominated by extensive flood basalt lavas with areal extents of $>100000$ km$^2$ and igneous volumes of $>100000$ km$^3$, most of which ($>75$\%) was expelled during relatively short periods ($\sim 1 - 5$ million years)~\cite{Bryan2008}. Furthermore, many LIPs have been associated with mantle plume activity and continental break-up~\cite{Ernst2010}. The Kap Washington Group volcanic sequence (North Greenland) and the Skaergaard intrusion (East Greenland) were considered for this search as parts of the High Arctic and North Atlantic LIPs, respectively~\cite{HALIP,NAIP}. Mid-ocean ridges, or rift volcanic zones where tectonic plates slowly move away from each other, are also of interest. Lava flows from Gakkel Ridge (Arctic Ocean)~\cite{GakkelRidge1,GakkelRidge2} provide attractive samples at very high latitude (84$^\circ$ N). Finally, some rock samples were selected on the basis that chemical analysis reveals hints of deep mantle origins. Some basaltic lavas from Coleman Nunatak (Marie Byrd Land, Antarctica) contain particularly high $^{206}$Pb/$^{204}$Pb ratios (denoted as high  $\mu$, or HIMU), which indicates low extent of melting and relatively deep origin~\cite{Coleman}. In addition, some of the lavas carry nodules of lherzolite, which have been carried up from the mantle source rocks without melting. Control samples, which should not contain stellar monopoles because they fail one of the search criteria, were also included: crust-derived lavas from a subduction zone (Antarctic Peninsula), and samples from a hotspot or mid-ocean ridge at low latitude (Hawaii, Mid-Atlantic Ridge and East Pacific Rise). The samples were shaped either as cylinders of 2.5~cm diameter and about 2.5~cm length, or crushed into fragments, which were placed into plastic cuboid boxes 2.3~cm on one side. The analysed samples are listed in Table~\ref{tab:samples} and amount to a total of 23.4~kg of search samples and 1.2~kg of control samples. 

\begin{widetext}

\begin{table}[t]\footnotesize
  \caption{\label{tab:samples} Characteristics of the rock samples used in this search. If not otherwise specified, they were emplaced during the Cenozoic era. Control samples are indicated with (c). The latitude corresponds to the location at the time of emplacement.}
  \begin{ruledtabular}
    \begin{tabular}{|l|c|l|l|c|c|}
      site                               & latitude     & tectonic setting          & rock type              & samples     & mass (kg) \\
      \hline 
      Iceland~\cite{iceland2}            & $64^\circ$ N & hotspot, mid-ocean ridge  & basalt                   & 144 & 5.916 \\
                                         &              &                           & gabbro                   &  26 & 1.404   \\
      Jan Mayen Island~\cite{JanMayen}   & $71^\circ$ N & hotspot                   & alkali basalt            &  6  & 0.139 \\
      Hawaii (c)                         & $21^\circ$ N & hotspot                   & tholeiitic basalt        &  17 & 0.610 \\  
      North Greenland~\cite{NG}          & $72^\circ$ N & LIP, 71-61 million        & alkali basalt, trachyte, &     &       \\
                                         &              & years old                 & trachyandesite, rhyolite &  73 & 1.779 \\
      East Greenland~\cite{Skaergaard}   & $68^\circ$ N & LIP, intrusion            & gabbro                   &  39 & 1.830 \\  
      Gakkel Ridge                       & $84^\circ$ N & mid-ocean ridge           & tholeiitic basalt        &  26 & 0.707 \\  
      Mid-Atlantic Ridge (c)             & $33^\circ$ S & mid-ocean ridge           & tholeiitic basalt        &   8 & 0.207 \\  
      East Pacific Rise (c)              & $28^\circ$ S & mid-ocean ridge           & tholeiitic basalt        &   7 & 0.241 \\  
      South. Victoria Land               & $77^\circ$ S & hotspot                   & basalt, basanite         & 233 & 8.163 \\      
      North. Victoria Land               & $72^\circ$ S & intraplate volcanism      & basalt, trachyte         &  12 & 0.335 \\      
      Marie Byrd Land~\cite{Coleman}     & $76^\circ$ S & intraplate volcanism      & alkali basalt (HIMU)     &  50 & 2.184 \\         
                                         &              &                           & lherzolite               &   3 & 0.148 \\         
                                         &              &                           & basalt, trachyte         &  17 & 0.440 \\      
      Ellsworth Land                     & $74^\circ$ S & intraplate volcanism      & basalt                   &  11 & 0.300 \\      
      Horlick Mountains                  & $87^\circ$ S & intraplate volcanism      & basalt                   &   1 & 0.021 \\      
      Antarctic Peninsula (c)            & $63^\circ$ S & subduction zone           & basalt                   &   5 & 0.146 \\  
      \hline 
      Total search                       &              &                           &                          & 641 & 23.366 \\    
      Total control (c)                  &              &                           &                          &  37 & 1.204 \\
    \end{tabular}
  \end{ruledtabular}
\end{table}

\end{widetext}

Samples were measured with a 2G Enterprises, model 755R, 3-axis DC-SQUID rock magnetometer housed in a shielded room at the Laboratory of Natural Magnetism, ETH Zurich. For magnetic dipoles the current reverts to zero on complete passage through the magnetometer superconducting coils. However, a monopole would leave the signature of a persistent current. This technique allows us to directly measure the magnetic charge contained inside a sample without the need to extract monopoles and with no mass dependence. Current measurements were performed in steps, including measurements where the sample is inside the sensing coils as well as 50~cm away from the sensing coils before and after the pass. Occasional passes with an empty sample holder were made for background subtraction. The persistent current is defined as the measured value after pass minus the value before pass (subtracting the same quantity for the empty holder), normalised such as to give the strength of magnetic pole contained in the sample in units of $g_D$. As described in detail in~\cite{LHCSQUID12}, calibration was performed using the convolution method, which consists of profiling the magnetometer response as a function of distance for a sample with well-known magnetisation and inferring the response for a monopole. As a calibration cross-check, the response to a magnetic pole was tested by introducing one extremity of a thin solenoid of 25~cm length with applied currents corresponding to values of magnetic charge of 0.124~$g_D$, 1.24~$g_D$, 12.4~$g_D$ and 124~$g_D$. The two methods yield consistent results within a normalisation uncertainty of 10\%.  

Samples with a total magnetisation $\geq 1.5\cdot 10^5~g_D$ (or magnetic dipole moment $\geq 4.4\cdot 10^{-5}$~Am$^2$) were found to sometimes cause the flux-locked loop of the SQUID to be lost and recovered at a different quantum level. This leaves a signal similar to what is expected from a monopole. Weaker moments generally did not show this effect. Precautions were therefore taken so that all samples would have magnetisation levels below $1.5\cdot 10^5~g_D$. Crushing the sample material into a gravel- or sand-sized powder randomises the magnetic moments from the constituent ferromagnetic minerals, which reduces the dipole signal. This method was frequently used in this study. Alternatively, the magnetisation can be reduced by more than an order of magnitude by exposing the sample to an alternating field. There is no risk of dislodging a trapped monopole if a binding energy of 100 keV or more is assumed. Demagnetisation was carried out only on 10\% of the Antarctic samples probed in this study.


\begin{figure}[tb]
  \begin{center}
    \includegraphics[width=1.\linewidth]{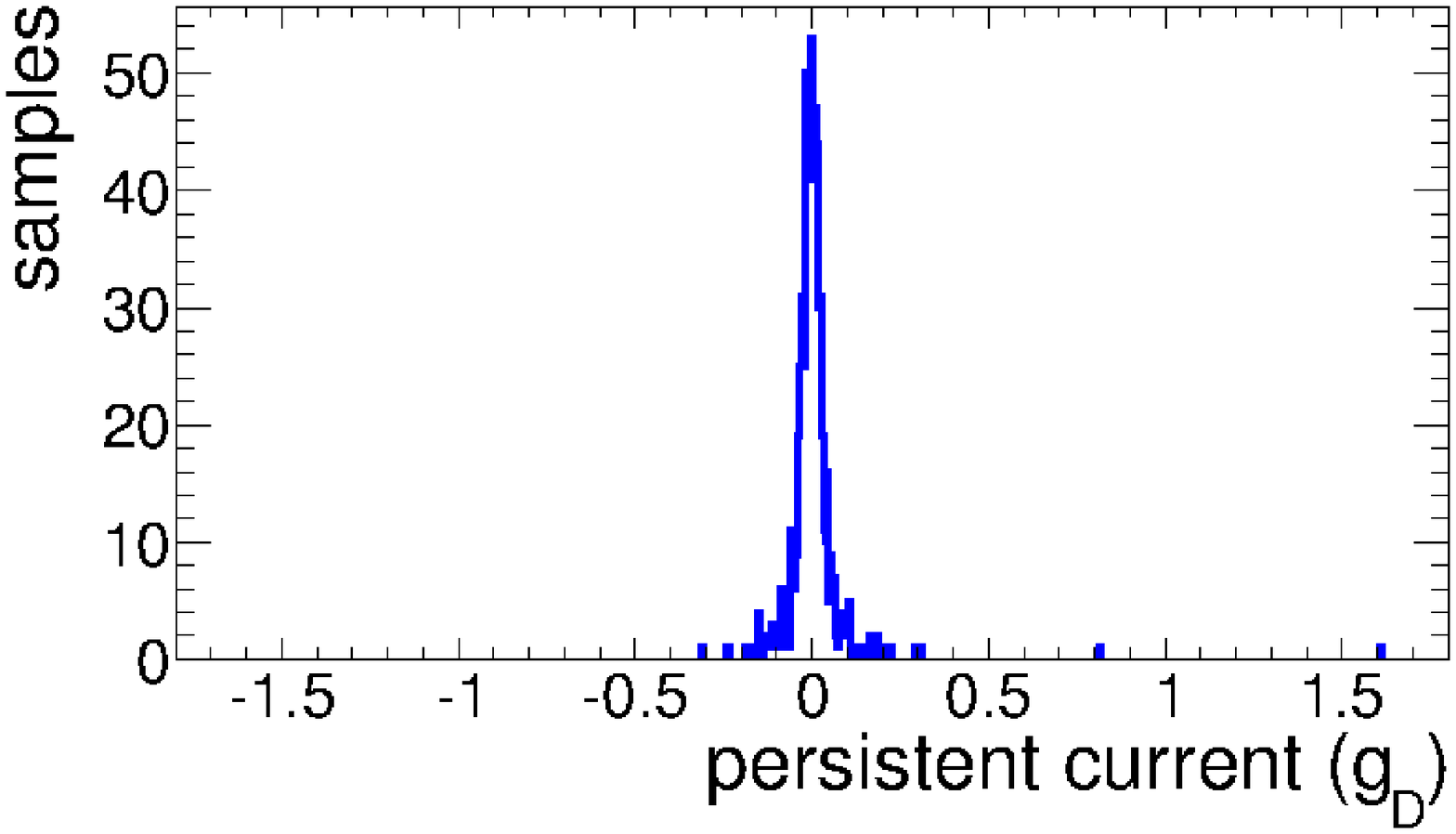}  
    \includegraphics[width=1.\linewidth]{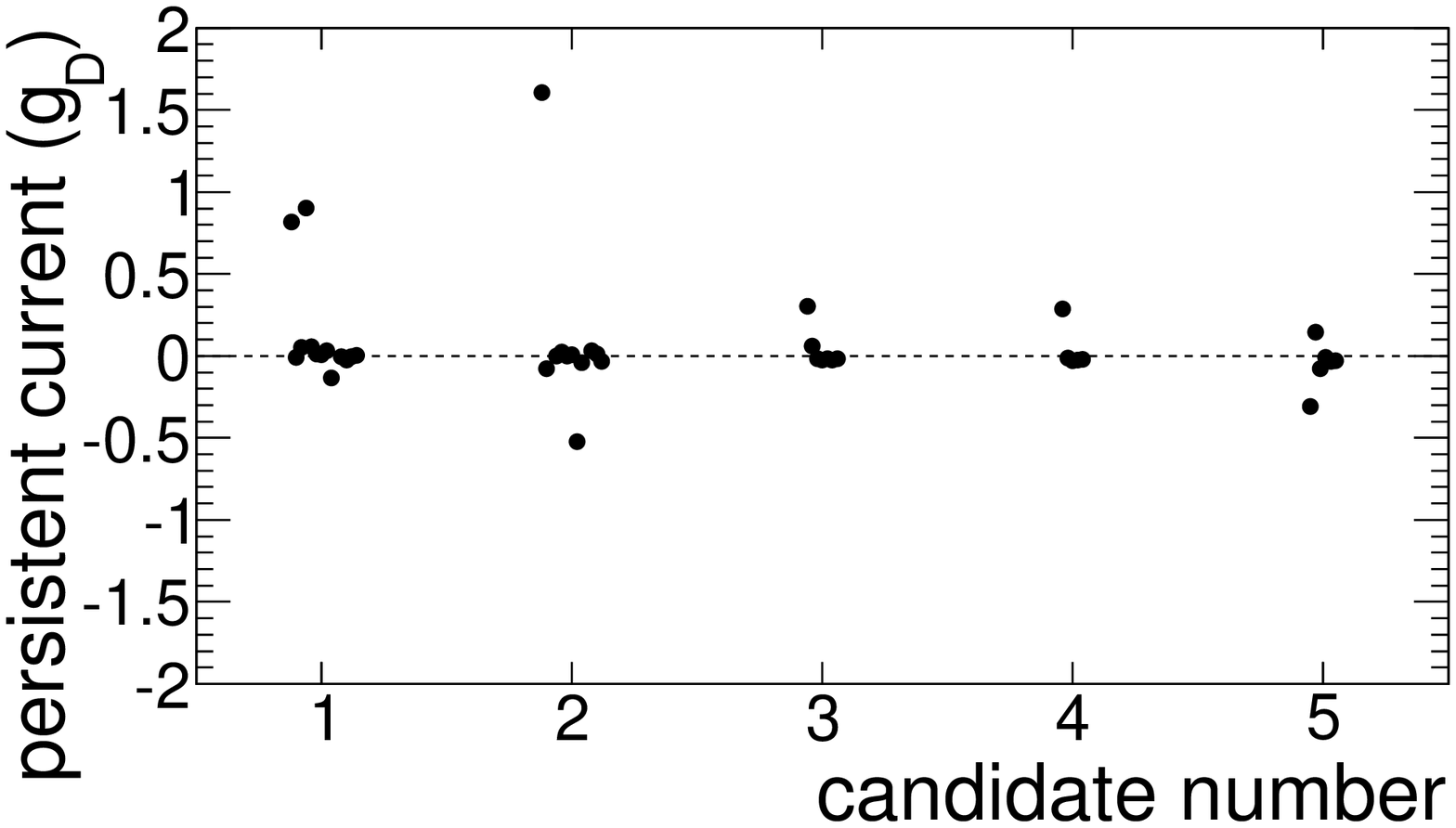}  
  \end{center}
  \caption{\label{fig:results} Top: persistent current after first passage through the magnetometer for all samples. Bottom: results of repeated measurements of candidate samples with absolute measured values in excess of $0.25~g_D$. }
\end{figure}


Measurements of persistent currents after first passage through the magnetometer are shown for all samples in Fig.~\ref{fig:results} (top). In the range from $-0.1$ to $0.1~g_D$, the distribution is Gaussian with mean value $-0.002\pm 0.001~g_D$ and standard deviation $0.026\pm 0.001~g_D$. Non-Gaussian tails slightly extend the distribution beyond this range. Five candidates out of 678 samples yield absolute values which deviate from zero by more than $0.25~g_D$. The two first of these candidates yield the largest values ($0.8~g_D$ and $1.6~g_D$) and also have total magnetisations in excess of $10^5~g_D$, close to the  $1.5\cdot 10^5~g_D$ limit beyond which measurements are known to be unreliable. Additional measurements of the five candidates using various orientations of the samples are shown in Fig.~\ref{fig:results} (bottom). These multiple measurements confirm the zero magnetic charge hypothesis. It is possible to get a rough estimate of the probability that a random sample containing a genuine monopole with $|g|=g_D$ would yield a persistent current close enough to zero to remain unnoticed. The probability to mismeasure the current by an absolute value which deviates from $g_D$ by less than $0.25~g_D$ is about $0.3\%$ (out of 678 samples, only the first candidate discussed above satisfies this condition, but some of the other candidates are close enough that we conservatively assume two). The probability to mismeasure the current in the direction where it would cancel out the current induced by a hypothetical monopole (whose charge can be positive or negative) is 1/2. Thus we obtain that $0.3\%/2=0.15\%$ of the signals with $|g|=g_D$ would escape detection; less if $|g|>g_D$. It is concluded that no monopoles with magnetic charge $|g|\geq g_D$ were present in the samples. 


The most extensive meteorite search to date -- the only other direct search with a non-negligible sensitivity to stellar monopoles -- sets a limit on the monopole density in meteoritic material of less than 
$2.1\cdot 10^{-5}$/gram at 90\% confidence level. The study analysed 112 kg of meteorites~\cite{Longo95}, among which $\sim 100$ kg are chondrites and can thus be assumed to consist of undifferentiated material from the primary solar nebula. This represents a little more than 4 times more material than used in the present search. As discussed above, for monopole mass and charge satisfying Equation~\ref{equ:forcebalance} for a position above the core-mantle boundary, this difference can be compensated for by an increase in monopole concentration of roughly a factor 6 in polar mantle-derived rocks due to monopole accumulation along the Earth's magnetic axis. One can think of two ways in which these results on stellar monopoles could be further improved in the future: by probing large ($>100$ kg) amounts of meteorites and polar rocks with a high-efficiency magnetometer, or by gaining access to new types of samples such as asteroid and comet fragments. 





In summary, massive monopoles of stellar origins would be absent from planetary surfaces and would tend to accumulate along the magnetic axis in planets with internal magnetic fields. If monopoles in the mass range $10^3\apprle m\apprle 10^{13}$ GeV are present within the Earth, they would be expected to be found inside the Earth's mantle below the geomagnetic poles. Assuming that monopoles bind strongly to nuclei, they would be trapped in mantle-derived rocks. This paper presents the first search for monopoles in polar igneous rocks. The search probed 23.4 kg of samples, for which a limit on the monopole density of
$9.8\cdot 10^{-5}$/gram at 90\% confidence level is set, which in a simple model translates into a limit of 
$1.6\cdot 10^{-5}$/gram in the matter averaged over the whole Earth. This search has a comparable or better sensitivity than the most extensive meteorite search and provides a novel probe of stellar monopoles in the Solar System. 


We are indebted to W.~E.~LeMasurier 
for providing rock samples from Coleman Nunatak, to ‎R.~G.~Tr\o nnes 
for providing a sample from the Beerenberg volcano, and to A.~Kontny for providing us samples from Hawaii and H.~B.~Mattsson for additional samples from Iceland. This research extensively used samples loaned from the United States Polar Rock Repository, which is sponsored by the United States National Science Foundation, Office of Polar Programs. This work was supported by a fellowship from the Swiss National Science Foundation and a grant from the Ernst and Lucie Schmidheiny Foundation.

%

\bibliography{SQUIDRocks12}

\end{document}